# Uncertainties of Parton Distribution Functions and Their Implications on Physical Predictions[†]

R. Brock, D. Casey, J. Huston, J. Kalk, J. Pumplin, D. Stump, W.K. Tung

Department of Physics and Astronomy
Michigan State University
East Lansing, MI 48824

We describe preliminary results from an effort to quantify the uncertainties in parton distribution functions and the resulting uncertainties in predicted physical quantities. The production cross section of the $W$ boson is given as a first example. Constraints due to the full data sets of the CTEQ global analysis are used in this study. Two complementary approaches, based on the Hessian and the Lagrange multiplier method respectively, are outlined. We discuss issues on obtaining meaningful uncertainty estimates that include the effect of correlated experimental systematic uncertainties and illustrate them with detailed calculations using one set of precision DIS data.

# 1  INTRODUCTION

Many measurements at the Tevatron rely on parton distribution functions (PDFs) for significant portions of their data analysis as well as the interpretation of their results. For example, in cross section measurements the acceptance calculation often relies on Monte Carlo (MC) estimates of the fraction of unobserved events. As another example, the measurement of the mass of the $W$ boson depends on PDFs via the modeling of the production of the vector boson in MC. In such cases, uncertainties in the PDFs contribute, by necessity, to uncertainties on the measured quantities. Critical comparisons between experimental data and the underlying theory are often even more dependent upon the uncertainties in PDFs. The uncertainties on the production cross sections for $W$ and $Z$ bosons, currently limited by the uncertainty on the measured luminosity, are approximately 4%. At this precision, any comparison with the theoretical prediction inevitably raises the question: How "certain" is the prediction itself?

A recent example of the importance of PDF uncertainty is the proper interpretation of the measurement of the high-$E_T$ jet cross-section at the Tevatron. When the first CDF measurement was published [1], there was a great deal of controversy over whether the observed excess, compared to theory, could be explained by deviations of the PDFs, especially the gluon, from the conventionally assumed behavior, or could it be the first signal for some new physics [2].

With the unprecedented precision and reach of many of the Run I measurements, understanding the implications of uncertainties in the PDFs has become a burning issue. During Run II (and later at LHC) this issue may strongly affect the uncertainty estimates in precision Standard Model studies, such as the all important $W$-mass measurement, as well as the signal and background estimates in searches for new physics.

In principle, it is the uncertainties on physical quantities due to parton distributions, rather than on the PDFs themselves, that is of primary concern. The latter are theoretical constructs which depend on the renormalization and factorization schemes; and there are strong correlations between PDFs of different flavors and from different values of $x$, which can compensate each other in the convolution integrals that relate them to physical cross-sections. On the other hand, since PDFs are universal, if we can obtain meaningful estimates of their uncertainties based on analysis of existing data, then the results can be applied to all processes that are of interest in the future. [3, 4]

One can attempt to assess directly the uncertainty on a specific physical prediction due to the full range of PDFs allowed by available experimental constraints. This approach will provide a more reliable estimate for the range of possible predictions for the physical variable under study, and may be the best course of action for ultra-precise measurements such as the mass of the $W$ boson or the $W$ production cross-section. However, such results are process-specific and therefore the analysis must be carried out for each case individually.

Until recently, the attempts to quantify either the uncertainties on the PDFs themselves (via uncertainties on their functional parameters, for instance) or the uncertainty on derived quantities due to variations in the PDFs have been rather unsatisfactory. Two commonly used methods are: (1) Comparing the predictions obtained with different PDF sets, *e.g.*, various CTEQ [5], MRS [6] and GRV [7] sets; (2) Within a given global analysis effort, varying individual functional parameters *ad hoc*, within limits considered to be consistent with the existing data, e.g. [8]. Neither method provides a systematic, quantitative measure of the uncertainties of the PDFs or their predictions. As a case in point, Fig. 1 shows how the calculated value of the cross section for $W$ boson production at the Tevatron varies with a set of historical CTEQ PDFs as well as the most recent CTEQ [5] and MRST [6] sets. Also shown are the most recent measurements from DØ and CDF[1].

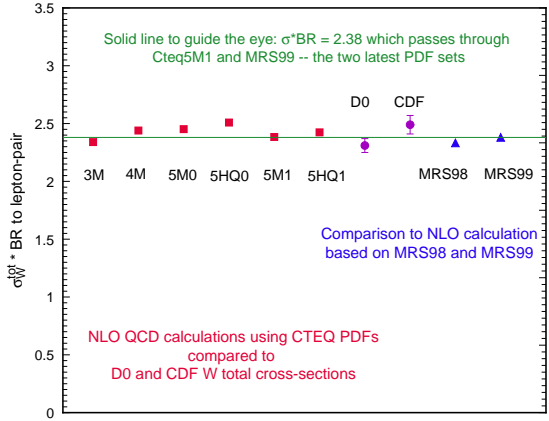

Figure 1: Predicted cross section for $W$ boson production for various PDFs.

While it is comforting to see that the predictions have remained within a narrow range, the variation observed cannot be characterized as a meaningful estimate of the uncertainty: (i) the variation with time reflects mostly the changes in experimental input to, or analysis procedure of, the global analyses; and (ii) the perfect agreement between the values of the most recent CTEQ5M1[2] and MRS99 sets must be fortuitous, since each group has also obtained other satisfactory sets which give rise to much larger variations of the $W$ cross section. The MRST group, in particular has examined the range of this variation by setting a variety of parameters to some extreme values [8]. These studies are useful but can not be considered quantitative or definitive. What is needed are methods that explore thoroughly the possible variations of the parton distribution functions.

It is important to recognize all potential **sources of uncertainty** in the determination of PDFs. Focusing on some of these, while neglecting significant others, may not yield practically useful results. Sources of uncertainty are listed below:

- **Statistical uncertainties** of the experimental data used to determine the PDFs. These vary over a wide range among the experiments used in a global analysis, but are straightforward to treat.

- **Systematic uncertainties** within each data set. There are typically many sources of experimental systematic uncertainty, some of which are highly correlated. These uncertainties can be treated by standard methods of probability theory *provided* they are precisely known, which unfortunately is often not the case – either because they may not be randomly distributed and/or because their estimation in practice involves subjective judgements.

- **Theoretical uncertainties** arising from higher-order PQCD corrections, resummation corrections near the boundaries of phase space, power-law (higher twist) and nuclear target corrections,

---

[1] It is interesting to note that much of the difference between the DØ and CDF $W$ cross sections is due to the different values of the total $p\bar{p}$ cross sections used

[2] CTEQ5M1 is an updated version of CTEQ5M differing only in a slight improvement in the QCD evolution (cf. note added in proof of [5]). The differences are completely insignificant for our purposes. Henceforth, we shall refer to them generically as CTEQ5M. Both sets can be obtained from the web address http://cteq.org/.

etc.

- Uncertainties due to the **parametrization of the non-perturbative PDFs**, $f_a(x, Q_0)$, at some low momentum scale $Q_0$. The specific choice of the functional form used at $Q_0$ introduces implicit correlations between the various $x$-ranges, which could be as important, if not more so, than the experimental correlations in the determination of $f_a(x, Q)$ for all $Q$.

Since strict quantitative statistical methods are based on idealized assumptions, such as random measurement uncertainties, an important trade-off must be faced in devising a **strategy** for the analysis of PDF uncertainties. If emphasis is put on the "rigor" of the statistical method, then many important experiments cannot be included in the analysis, either because the published errors appear to fail strict statistical tests or because data from different experiments appear to be mutually exclusive in the parton distribution parameter space [4]. If priority is placed on using the maximal experimental constraints from available data, then standard statistical methods may not apply, but must be supplemented by physical considerations, taking into account experimental and theoretical limitations. We choose the latter tack, pursuing the determination of the uncertainties in the context of the current CTEQ global analysis. In particular, we include the same body of the world's data as constraints in our uncertainty study as that used in the CTEQ5 analysis; and adopt the "best fit" – the CTEQ5M1 set – as the base set around which the uncertainty studies are performed. In practice, there are unavoidable choices (and compromises) that must be made in the analysis. (Similar subjective judgements often are also necessary in estimating certain systematic errors in experimental analyses.) The most important consideration is that quantitative results must remain robust with respect to reasonable variations in these choices.

In this Report we describe preliminary results obtained by our group using the two approaches mentioned earlier. In Section 3 we focus on the error matrix, which characterizes the general uncertainties of the non-perturbative PDF parameters. In Sections 4 and 5 we study specifically the production cross section $\sigma_W$ for $W^\pm$ bosons at the Tevatron, to estimate the uncertainty of the prediction of $\sigma_W$ due to PDF uncertainty. We start in Section 2 with a review of some aspects of the CTEQ global analysis on which this study is based.

## 2 Elements of the Base Global Analysis

Since our strategy is based on using the existing framework of the CTEQ global analysis, it is useful to review some of its features pertinent to the current study [5].

**Data selection:** Table 1 shows the experimental data sets included in the CTEQ5 global analysis, and in the current study. For neutral current DIS data only the most accurate proton and deuteron target measurements are kept, since they are the "cleanest" and they are already extremely extensive. For charged current (neutrino) DIS data, the significant ones all involve a heavy (Fe) target. Since these data are crucial for the determination of the normalization of the gluon distribution (indirectly via the momentum sum rule), and for quark flavor differentiation (in conjunction with the neutral current data), they play an important role in any comprehensive global analysis. For this purpose, a heavy-target correction is applied to the data, based on measured ratios for heavy-to-light targets from NMC and other experiments. Direct photon production data are not included because of serious theoretical uncertainties, as well as possible inconsistencies between existing experiments. Cf. [5] and [9]. The combination of neutral and charged DIS, lepton-pair production, lepton charge asymmetry, and inclusive large-$p_T$ jet production processes provides a fairly tightly constrained system for the global analysis of PDFs. In total, there are ∼1300 data points which meet the minimum momentum scale cuts which must be imposed to ensure that PQCD applies. The fractional uncertainties on these points are distributed roughly like $dF/F$ over the range $F = 0.003 - 0.4$.

| Process | Experiment | Measurable | $N_{data}$ |
|---|---|---|---|
| DIS | BCDMS[10] | $F_{2\,H}^{\mu}, F_{2\,D}^{\mu}$ | 324 |
| | NMC [11] | $F_{2\,H}^{\mu}, F_{2\,D}^{\mu}$ | 240 |
| | H1 [12] | $F_{2\,H}^{e}$ | 172 |
| | ZEUS[13] | $F_{2\,H}^{e}$ | 186 |
| | CCFR [14] | $F_{2\,Fe}^{\nu}, x\,F_{3\,Fe}^{\nu}$ | 174 |
| Drell-Yan | E605[15] | $s d\sigma/d\sqrt{\tau}dy$ | 119 |
| | E866 [16] | $\sigma(pd)/2\sigma(pp)$ | 11 |
| | NA-51[17] | $A_{DY}$ | 1 |
| W-prod. | CDF [18] | Lepton asym. | 11 |
| Incl. Jet | CDF [19] | $d\sigma/dE_t$ | 33 |
| | D0[20] | $d\sigma/dE_t$ | 24 |

Table 1: List of processes and experiments used in the CTEQ5M Global analysis. The total number of data points is 1295.

**Parametrization:**  The non-perturbative parton distribution functions $f_a(x,Q)$ at a low momentum scale $Q = Q_0$ are parametrized by a set of functions of $x$, corresponding to the various flavors $a$. For this analysis, $Q_0$ is taken to be 1 GeV. The specific functional forms and the choice of $Q_0$ are not important, as long as the parametrization is general enough to accommodate the behavior of the true (but unknown) non-perturbative PDFs. The CTEQ analysis adopts the functional form

$$a_0 x^{a_1}(1-x)^{a_2}(1+a_3 x^{a_4}).$$

for most quark flavors as well as for the gluon.[3] After momentum and quark number sum rules are enforced, there are 18 free parameters left over, hereafter referred to as "shape parameters" $\{a_i\}$. The PDFs at $Q > Q_0$ are determined from $f_a(x,Q_0)$ by evolution equations from the renormalization group.

**Fitting:**  The values of $\{a_i\}$ are determined by fitting the global experimental data to the theoretical expressions which depend on these parameters. The fitting is done by minimizing a global "chi-square" function, $\chi^2_{\text{global}}$. The quotation mark indicates that this function serves as a *figure of merit* of the quality of the global fit; it does not necessarily have the full significance associated with rigorous statistical analysis, for reasons to be discussed extensively throughout the rest of this report. In practice, this function is defined as:

$$\begin{aligned}\chi^2_{\text{global}} &= \sum_n \sum_i w_n \left[ (N_n d_{ni} - t_{ni}) / \sigma^d_{ni} \right]^2 \\ &+ \sum_n \left[ (1 - N_n) / \sigma^N_n \right]^2 \end{aligned} \quad (1)$$

where $d_{ni}$, $\sigma^d_{ni}$, and $t_{ni}$ denote the data, measurement uncertainty, and theoretical value (dependent on $\{a_i\}$) for the $i^{\text{th}}$ data point in the $n^{\text{th}}$ experiment. The second term allows the absolute normalization ($N_n$) for each experiment to vary, constrained by the published normalization uncertainty ($\sigma^N_n$). The $w_n$ factors are weights applied to some critical experiments with very few data points, which are known (from physics considerations) to provide useful constraints on certain unique features of PDFs not afforded by other experiments. Experience shows that without some judiciously chosen weights, these experimental data points will have no influence in the global fitting process. The use of these weighing factors, to enable the relevant unique constraints, amounts to imposing certain prior probability (based on physics knowledge) to the statistical analysis.

---

[3] An exception is that recent data from E866 seem to require the ratio $\bar{d}/\bar{u}$ to take a more unconventional functional form.

In the above form, $\chi^2_{\text{global}}$ includes for each data point the random statistical uncertainties and the combined systematic uncertainties in uncorrelated form, as presented by most experiments in the published papers. These two uncertainties are *combined in quadrature* to form $\sigma^d_{ni}$ in Eq. 1. Detailed point to point correlated systematic uncertainties are not available in the literature in general; however, in some cases, they can be obtained from the experimental groups. For global fitting, uniformity in procedure with respect to all experiments favors the usual practice of merging them into the uncorrelated uncertainties. For the study of PDF uncertainties, we shall discuss this issue in more detail in Section 5.

**Goodness-of-fit for CTEQ5M:** Without going into details, Fig. 2 gives an overview of how well CTEQ5m fits the total data set. The graph is a histogram of the variable $x \equiv (d - t)/\sigma$ where $d$ is a data value, $\sigma$ the uncertainty of that measurement (statistical and systematic combined), and $t$ the theoretical value for CTEQ5m. The curve in Fig. 2 has no adjustable parameters; it is the Gaussian with width 1 normalized to the total number of data points (1295). Over the entire data set, the theory fits the data within the assigned uncertainties $\sigma^d_{ni}$, indicating that those uncertainties are numerically consistent with the actual measurement fluctuations. Similar histograms for the individual experiments reveal various deviations from the theory, but *globally* the data have a reasonable Gaussian distribution around CTEQ5M.

## 3   Uncertainties on PDF parameters: The Error Matrix

We now describe results from an investigation of the behavior of the $\chi^2_{\text{global}}$ function at its minimum, using the standard error matrix approach [21]. This allows us to determine which combinations of parameters are contributing the most to the uncertainty.

At the minimum of $\chi^2_{\text{global}}$, the first derivatives with respect to the $\{a_i\}$ are zero; so near the minimum, $\chi^2_{\text{global}}$ can be approximated by

$$\chi^2_{\text{global}} = \chi^2_0 + \frac{1}{2}\sum_{i,j} F_{ij} y_i y_j \qquad (2)$$

where $y_i = a_i - a_{0i}$ is the displacement from the minimum, and $F_{ij}$ is the *Hessian*, the matrix of second derivatives. It is natural to define a new set of coordinates using the complete orthonormal set of eigenvectors of the symmetric matrix $F_{ij}$ as basis vectors. These vectors can be ordered by their eigenvalues $e_i$. Each eigenvalue is a quantitative measure of the uncertainties in the shape parameters $\{a_i\}$ for displacements in parameter space in the direction of the corresponding eigenvector. The quantity $\ell_i \equiv 1/\sqrt{e_i}$ is the distance in the 18 dimensional parameter space, in the direction of eigenvector $i$, that makes a unit increase in $\chi^2_{\text{global}}$. If the only measurement uncertainty were uncorrelated gaussian uncertainties, then $\ell_i$ would be one standard deviation from the best fit in the direction of the eigenvector. The inverse of the Hessian is the error matrix.

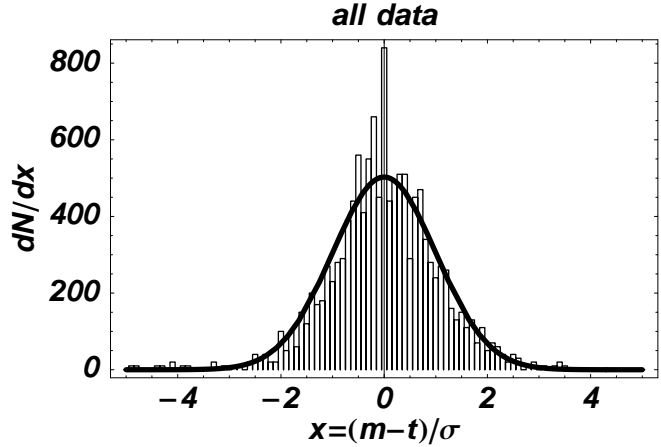

Figure 2: Histogram of the $(measurement - theory)$ for all data points in the CTEQ5m fit.

Because the real uncertainties, for the wide variety of experiments included, are far more complicated than assumed in the ideal situation, the quantitative measure of a given increase in $\chi^2_{global}$ carries little true statistical meaning. However, qualitatively, the Hessian gives an analytic picture of $\chi^2_{\text{global}}$ near its minimum in $\{a_i\}$ space, and hence allows us to identify the particular degrees of freedom that need further experimental input in future global analyses.

From calculations of the Hessian we find that the eigenvalues vary over a wide range. Figure 3 shows a graph of the eigenvalues of $F_{ij}$, on a logarithmic scale. The vertical axis is $\ell_i = 1/\sqrt{e_i}$, the distance of a "standard deviation" along the $i^{\text{th}}$ eigenvector. These distances range over 3 orders of magnitude. Large eigenvalues of $F_{ij}$ correspond to "steep directions" of $\chi^2_{\text{global}}$. The corresponding eigenvectors are combinations of shape parameters that are well determined by current data. For example, parameters that govern the valence $u$ and $d$ quarks at moderate $x$ are sharply constrained by DIS data. Small eigenvalues of $F_{ij}$ correspond to "flat directions" of $\chi^2_{\text{global}}$. In the directions of these eigenvectors, $\chi^2_{\text{global}}$ changes little over large distances in $\{a_i\}$ space. For example, parameters that govern the large-$x$ behavior of the gluon distribution, or differences between sea quarks, properties of the nucleon that are not accurately determined by current data, contribute to the flat directions. The existence of flat directions is inevitable in global fitting, because as the data improve it only makes sense to maintain enough flexibility for $f_a(x, Q_0)$ to fit the available experimental constraints.

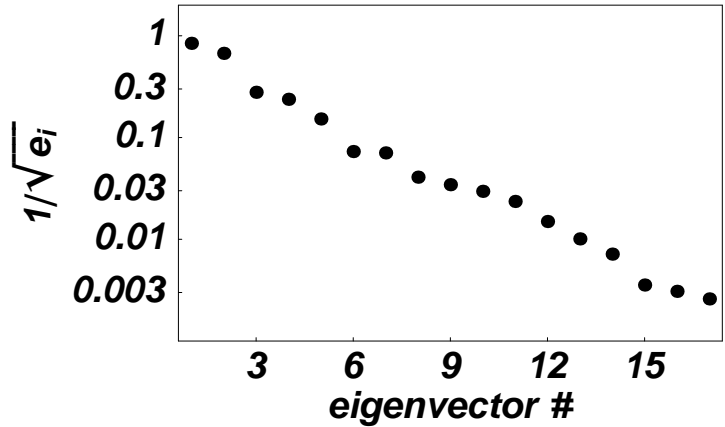

Figure 3: Plot of the eigenvalues of the Hessian. The vertical axis is $\ell_i = 1/\sqrt{e_i}$.

Because the eigenvalues of the Hessian have a large range of values, efficient calculation of $F_{ij}$ requires an adaptive algorithm. In principle $F_{ij}$ is the matrix of second derivatives at the minimum of $\chi^2_{\text{global}}$, which could be calculated from very small finite differences. In practice, small computational errors in the evaluation of $\chi^2_{\text{global}}$ preclude the use of a very small step size. Coarse grained finite differences yield a more accurate calculation of the second derivatives. But because the variation of $\chi^2_{\text{global}}$ varies markedly in different directions, it is important to use a grid in $\{a_i\}$ space with small steps in steep directions and large steps in flat directions. This grid is generated by an iterative procedure, in which $F_{ij}$ converges to a good estimate of the second derivatives.

From calculations of $F_{ij}$ we find that the minimum of $\chi^2_{\text{global}}$ is fairly quadratic over large distances in the parameter space. Figures 4 and 5 show the behavior of $\chi^2_{\text{global}}$ near the minimum along each of the 18 eigenvectors. $\chi^2_{\text{global}}$ is plotted on the vertical axis, and the variable on the horizontal axis is the distance in $\{a_i\}$ space in the direction of the eigenvector, in units of $\ell_i = 1/\sqrt{e_i}$. There is some nonlinearity, but it is small enough that the Hessian can be used as an analytic model of the functional dependence of $\chi^2_{\text{global}}$ on the shape parameters.

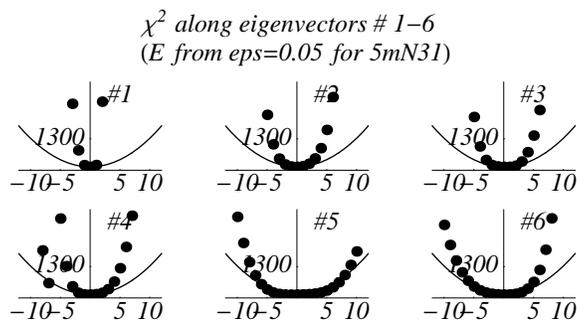

Figure 4: Value of $\chi^2$ along the six eigenvectors with the largest eigenvalues.

In a future paper we will provide details on the uncertainties of the original shape parameters

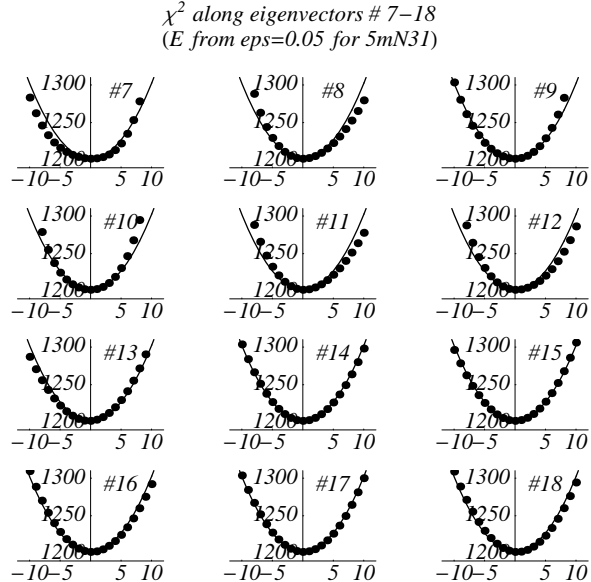

*$\chi^2$ along eigenvectors # 7–18*
*(E from eps=0.05 for 5mN31)*

Figure 5: Value of $\chi^2$ along the 12 eigenvectors with the smallest eigenvalues.

$\{a_i\}$. But it should be remembered that these parameters specify the PDFs at the low $Q$ scale, and applications of PDFs to Tevatron experiments use PDFs at a high $Q$ scale. The evolution equations determine $f(x, Q)$ from $f(x, Q_0)$, so the functional form at $Q$ depends on the $\{a_i\}$ in a complicated way.

## 4 Uncertainty on $\sigma_W$: the Lagrange Multiplier Method

In this Section, we determine the variation of $\chi^2_{\text{global}}$ as a function of a single measurable quantity. We use the production cross section for $W$ bosons ($\sigma_W$) as an archetype example. The same method can be applied to any other physical observable of interest, for instance the Higgs production cross section, or to certain measured differential distributions. The aim is to quantify the uncertainty on that physical observable due to uncertainties of the PDFs integrated over the entire PDF parameter space.

Again, we use the standard CTEQ5 analysis tools and results [5] as the starting point. The "best fit" is the CTEQ5M1 set. A natural way to find the limits of a physical quantity $X$, such as $\sigma_W$ at $\sqrt{s} = 1.8\,\text{TeV}$, is to take $X$ as one of the search parameters in the global fit and study the dependence of $\chi^2_{\text{global}}$ for the 15 base experimental data sets on $X$.

Conceptually, we can think of the function $\chi^2_{\text{global}}$ that is minimized in the fit as a function of $a_1, \ldots, a_{17}, X$ instead of $a_1, \ldots, a_{18}$. This idea could be implemented directly in principle, but a more convenient way to do the same thing in practice is through Lagrange's method of undetermined multipliers. One minimizes, with respect to the $\{a_i\}$, the quantity

$$F(\lambda) = \chi^2_{\text{global}} + \lambda X(a_1, \ldots, a_{18}) \qquad (3)$$

for a fixed value of $\lambda$, the Lagrange multiplier. By minimizing $F(\lambda)$ for many values of $\lambda$, we map out $\chi^2_{\text{global}}$ as a function of $X$. The minimum of $F$ for a given value of $\lambda$ is the best fit to the data for the corresponding value of $X$, *i.e.*, evaluated at the minimum.

Figure 6 shows $\chi^2_{\text{global}}$ for the 15 base experimental data sets as a function of $\sigma_W$ at the Tevatron. The horizontal axis is $\sigma_W$ times the branching ratio for $W \to$ leptons, in nb. The CTEQ5m prediction is $\sigma_W \cdot BR_{\text{lep}} = 2.374$ nb. The vertical dashed lines are $\pm 3\%$ and $\pm 5\%$ deviations from the CTEQ5m prediction.

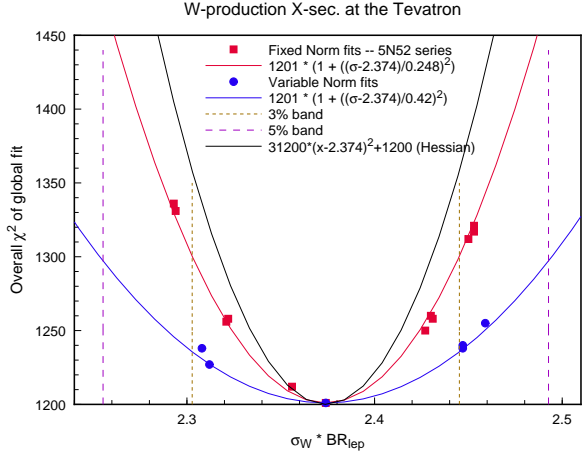

Figure 6: $\chi^2$ of the base experimental data sets *versus* $\sigma_W \cdot BR_{\text{lep}}$, the $W$ production cross-section at the Tevatron times lepton branching ratio, in nb.

The two parabolas associated with points in Fig. 6 correspond to different treatments of the normalization factor $N_n$ in Eq. 1. The dots are variable norm fits, in which $N_n$ is allowed to float, taking into account the experimental normalization uncertainties, and $F(\lambda)$ is minimized with respect to $N_n$. The justification for this procedure is that overall normalization is a common systematic uncertainty. The square points are fixed norm fits, in which all $N_n$ are held fixed at their values for the global minimum (CTEQ5m). These two procedures represent extremes in the treatment of normalization uncertainty. The parabolas are just least-square fits to the points in the two cases.

The other curve in Fig. 6 was calculated using the Hessian method. The Hessian $F_{ij}$ is the matrix of second derivatives of $\chi^2_{\text{global}}$ with respect to the shape parameters $\{a_i\}$. The derivatives (first and second) of $\sigma_W$ may also be calculated by finite differences. Using the resultant quadratic approximations for $\chi^2_{\text{global}}(a)$ and $\sigma_W(a)$, one may minimize $\chi^2_{\text{global}}$ with $\sigma_W$ fixed. Since this calculation keeps the normalization factors constant, it should be compared with the fixed norm fits from the Lagrange multiplier method. The fact that the Hessian and Lagrange multiplier methods yield similar results lends support to both approaches; the small difference between them indicates that the quadratic functional approximations for $\chi^2_{\text{global}}$ and $\sigma_W$ are only approximations.

For the quantitative analysis of uncertainties, the important question is: How large an increase in $\chi^2_{\text{global}}$ should be taken to define the likely range of uncertainty in $X$? There is an elementary statistical theorem that states that $\Delta\chi^2 = 1$ in a constrained fit corresponds to 1 standard deviation of the constrained quantity $X$. However, the theorem relies on the assumption that the uncertainties are gaussian, uncorrelated, and correctly estimated in magnitude. Because these conditions do not hold for the full data set (of $\sim 1300$ points from 15 different experiments), this theorem cannot be naively applied quantitatively.[4] Indeed, it can be shown that, if the measurement uncertainties are correlated, and the correlation is not properly taken into account in the definition of $\chi^2_{\text{global}}$, then a standard deviation may vary over the entire range from $\Delta\chi^2 = 1$ to $\Delta\chi^2 = N$ (the total number of data points – $\sim 1300$ in our case).

# 5 STATISTICAL ANALYSIS WITH SYSTEMATIC UNCERTAINTIES

Fig. 6 shows how the fitting function $\chi^2_{\text{global}}$ increases from its minimum value, at the best global fit, as the cross-section $\sigma_W$ for $W$ production is forced away from the prediction of the global fit. The next step in our analysis of PDF uncertainty is to use that information, or some other analysis, to estimate the uncertainty in $\sigma_W$. In ideal circumstances we could say that a certain increase of $\chi^2_{\text{global}}$ from the minimum value, call it $\Delta\chi^2$, would correspond to a standard deviation of the global measurement uncertainty. Then a horizontal line on Fig. 6 at $\chi^2_{\text{min}} + \Delta\chi^2$ would indicate the probable range of $\sigma_W$, by the intersection with the parabola of $\chi^2_{\text{global}}$ versus $\sigma_W$.

However, such a simple estimate of the uncertainty of $\sigma_W$ is not possible, because the fitting function $\chi^2_{\text{global}}$ does not include the *correlations* between systematic uncertainties. The uncertainty $\sigma^d_{ni}$ in the definition (1) of $\chi^2_{\text{global}}$ combines *in quadrature* the statistical and systematic uncertainties for each data point; that is, it treats the systematic uncertainties as uncorrelated. The standard theorems of statistics for Gaussian probability distributions of random uncertainties do not apply to $\chi^2_{\text{global}}$.

Instead of using $\chi^2_{\text{global}}$ to estimate confidence levels on $\sigma_W$, we believe the best approach is to carry out a thorough statistical analysis, including the correlations of systematic uncertainties, on individual experiments used in the global fit for which detailed information is available. We will describe here such an analysis for the measurements of $F_2(x, Q)$ by the H1 experiment [12] at HERA, as a case study. In a future paper, we will present similar calculations for other experiments. The H1 experiment has provided a detailed table of measurement uncertainties – statistical and systematic – for their measurements of $F_2(x, Q)$. [12] The CTEQ program uses 172 data points from H1 (requiring the cut $Q^2 > 5\,\text{GeV}^2$). For each measurement $d_j$ (where $j = 1 \ldots 172$) there is a statistical uncertainty $\sigma_{0j}$, an uncorrelated systematic uncertainty $\sigma_{1j}$, and a set of 4 correlated systematic uncertainties $a_{jk}$ where $k = 1 \ldots 4$. (In fact there are 8 correlated uncertainties listed in the H1 table. These correspond to 4 pairs. Each pair consists of one standard deviation in the positive sense, and one standard deviation in the negative sense, of some experimental parameter. For this first analysis, we have approximated each pair of uncertainties by a single, symmetric combination, equal in magnitude to the average magnitude of the pair.)

To judge the uncertainty of $\sigma_W$, as constrained by the H1 data, we will compare the H1 data to the global fits in Fig. 6. The comparison is based on the true, statistical $\chi^2$, including the correlated

---

[4]It has been shown by Giele *et.al.* [4], that, taken literally, only one or two selected experiments satisfy the standard statistical tests.

| Lagrange multiplier | $\sigma_W \cdot B$ in nb | $\chi^2/172$ | probability |
|---:|---:|---:|---:|
| 3000 | 2.294 | 1.0847 | 0.212 |
| 2000 | 2.321 | 1.0048 | 0.468 |
| 1000 | 2.356 | 0.9676 | 0.605 |
| 0 | 2.374 | 0.9805 | 0.558 |
| -1000 | 2.407 | 1.0416 | 0.339 |
| -2000 | 2.431 | 1.0949 | 0.187 |
| -3000 | 2.450 | 1.1463 | 0.092 |

Table 2: Comparison of H1 data to the PDF fits with constrained values of $\sigma_W$.

uncertainties, which is given by

$$\chi^2 = \sum_j \frac{(d_j - t_j)^2}{\sigma_j^2} - \sum_{kk'} B_k \left(A^{-1}\right)_{kk'} B_{k'}. \tag{4}$$

The index $j$ labels the data points and runs from 1 to 172. The indices $k$ and $k'$ label the source of systematic uncertainty and run from 1 to 4. The combined uncorrelated uncertainty $\sigma_j$ is $\sqrt{\sigma_{0j}^2 + \sigma_{1j}^2}$. The second term in (4) comes from the correlated uncertainties. $B_k$ is the vector

$$B_k = \sum_j \frac{(d_j - t_j)\, a_{jk}}{\sigma_j^2}, \tag{5}$$

and $A_{kk'}$ is the matrix

$$A_{kk'} = \delta_{kk'} + \sum_j \frac{a_{jk} a_{jk'}}{\sigma_j^2}. \tag{6}$$

Assuming the published uncertainties $\sigma_{0j}$, $\sigma_{1j}$ and $a_{jk}$ accurately reflect the measurement fluctuations, $\chi^2$ would obey a chi-square distribution if the measurements were repeated many times. Therefore the chi-square distribution with 172 degrees of freedom provides a basis for calculating *confidence levels* for the global fits in Fig. 6.

Table 2 shows $\chi^2$ for the H1 data compared to seven of the PDF fits in Fig. 6. The center row of the Table is the global best fit – CTEQ5m. The other rows are fits obtained by the Lagrange multiplier method for different values of the Lagrange multiplier. The best fit to the H1 data, *i.e.*, the smallest $\chi^2$, is not CTEQ5m (the best global fit) but rather the fit with Lagrange multiplier 1000 for which $\sigma_W$ is 0.8% smaller than the prediction of CTEQ5m. Forcing the $W$ cross section values away from the prediction of CTEQ5m causes an increase in $\chi^2$ for the DIS data. At $\sqrt{s} = 1.8\,\text{TeV}$, $W$ production is mainly from $q\bar{q} \to W^+ W^-$ with moderate values of $x$ for $q$ and $\bar{q}$, *i.e.*, values in the range of DIS experiments. Forcing $\sigma_W$ higher (or lower) requires a higher (or lower) valence quark density in the proton, in conflict with the DIS data, so $\chi^2$ increases.

The final column in Table 2, labeled "probability", is computed from the chi-square distribution with 172 degrees of freedom. This quantity is the probability for $\chi^2$ to be greater than the value calculated from the existing data, if the H1 measurements were to be repeated. So, for example, the fit with Lagrange multiplier $-3000$, which corresponds to $\sigma_W$ being 3.2% larger than the CTEQ5m prediction, has probability 0.092. In other words, if the H1 measurements could be repeated many times, in only 9.2% of trials would $\chi^2$ be greater than or equal to the value that has been obtained with the existing data. This probability represents a confidence level for the value of $\sigma_W$ that was forced on the PDF by setting the Lagrange multiplier equal to -3000. At the 9.2% confidence level we can say that $\sigma_W \cdot BR_{\text{lep}}$ is less than 2.450 nb, based on the H1 data. Similarly, at the 21.2% confidence level we can say that $\sigma_W \cdot BR_{\text{lep}}$ is greater than 2.294 nb.

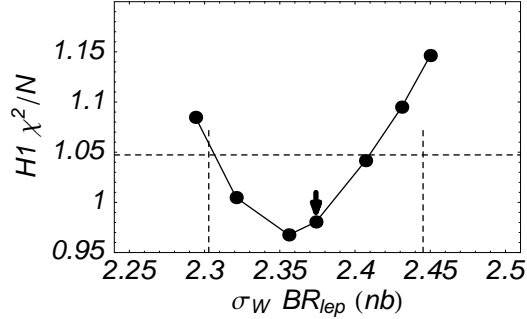

Figure 7: $\chi^2/N$ of the H1 data, including error correlations, compared to PDFs obtained by the Lagrange multiplier method for constrained values of $\sigma_W$.

Fig. 7 is a graph of $\chi^2/N$ for the H1 data compared to the PDF fits in Table 2. This figure may be compared to Fig. 6. The CTEQ5 prediction of the $W$ production cross-section is shown as an arrow, and the vertical dashed lines are $\pm 3\%$ away from the CTEQ5m prediction. The horizontal dashed line is the 68% confidence level on $\chi^2/N$ for $N = 172$ degrees of freedom. The comparison with H1 data alone indicates that the uncertainty on $\sigma_W$ is $\sim 3\%$.

There is much more to say about $\chi^2$ and confidence levels. In a future paper we will discuss statistical calculations for other experiments in the global data set. The H1 experiment is a good case, because for H1 we have detailed information about the correlated uncertainties. But it may be somewhat fortuitous that the $\chi^2$ per data point for CTEQ5m is so close to 1 for the H1 data set. In cases where $\chi^2/N$ is not close to 1, which can easily happen if the estimated systematic uncertainties are not textbook-like, we must supply further arguments about confidence levels. For experiments with many data points, like 172 for H1, the chi-square distribution is very narrow, so a small inaccuracy in the estimate of $\sigma_j$ may translate to a large uncertainty in the calculation of confidence levels based on the absolute value of $\chi^2$. Because the estimation of experimental uncertainties introduces some uncertainty in the value of $\chi^2$, it is not really the *absolute* value of $\chi^2$ that is important, but rather the *relative* value compared to the value at the global minimum. Therefore, we might study *ratios* of $\chi^2$'s to interpret the variation of $\chi^2$ with $\sigma_W$.

# 6 CONCLUSIONS

It has been widely recognized by the HEP community, and it has been emphasized at this workshop, that PDF phenomenology must progress from the past practice of periodic updating of *representa-* *tive* PDF sets to a systematic effort to map out the uncertainties, both on the PDFs themselves and on physical observables derived from them. For the analysis of PDF uncertainties, we have only addressed the issues related to the treatment of experimental uncertainties. Equally important for the ultimate goal, one must come to grips with uncertainties associated with theoretical approximations and phenomenological parametrizations. Both of these sources of uncertainties induce highly correlated uncertainties, and they can be numerically more important than experimental uncertainties in some cases. Only a balanced approach is likely to produce truly useful results. Thus, great deal of work lies ahead.

This report described first results from two methods for quantifying the uncertainty of parton distribution functions associated with experimental uncertainties. The specific work is carried out as extensions of the CTEQ5 global analysis. The same methods can be applied using other parton distributions as the starting point, or using a different parametrization of the non-perturbative PDFs. We have indeed tried a variety of such alternatives. The results are all similar to those presented above. The robustness of these results lends confidence to the general conclusions.

The Hessian, or error matrix method reveals the uncertainties of the shape parameters used in the functional parametrization. The behavior of $\chi^2_{\text{global}}$ in the neighborhood of the minimum is well described by the Hessian if the minimum is quadratic.

The Lagrange multiplier method produces constrained fits, *i.e.*, the best fits to the global data set for specified values of some observable. The increase of $\chi^2_{\text{global}}$, as the observable is forced away from the predicted value, indicates how well the current data on PDFs determines the observable. The constrained fits generated by the Lagrange multiplier method may be compared to data from individual experiments, taking into account the uncertainties in the data, to estimate confidence levels for the constrained variable. For example, we estimate that the uncertainty of $\sigma_W$ attributable to PDFs is $\pm 3\%$.

Further work is needed to apply these methods to other measurements, such as the $W$ mass or the forward-backward asymmetry of $W$ production in $p\bar{p}$ collisions. Such work will be important for analyzing high precision measurements involving the electroweak bosons in the Tevatron Run II era.

# Acknowledgement

This research is supported in part by the National Science Foundation.